# A new simulation algorithm for lattice QCD with dynamical quarks*

B. Bunk[a], K. Jansen[b], B. Jegerlehner[c], M. Lüscher[b], H. Simma[b] and R. Sommer[d]

[a]Humboldt Universität, Institut für Physik
Invalidenstrasse 110, D-10099 Berlin, Germany

[b]Deutsches Elektronen-Synchrotron DESY
Notkestrasse 85, D-22603 Hamburg, Germany

[c]Max-Planck-Institut für Physik
Föhringerring 6, D-80805 München, Germany

[d]CERN, Theory Division
CH-1211 Genève 23, Switzerland

A previously introduced multi-boson technique for the simulation of QCD with dynamical quarks is described and some results of first test runs on a $6^3 \times 12$ lattice with Wilson quarks and gauge group SU(2) are reported.

## 1. INTRODUCTION

The basic idea of the algorithm proposed in ref.[1] is to map lattice QCD to a *local* bosonic theory and to simulate the latter using standard techniques. The method is generally applicable, but attention is here restricted to the case of two degenerate flavours of Wilson quarks on a four-dimensional lattice with periodic boundary conditions in all directions. We shall first recall the basic steps leading to the bosonic theory and then present some results from test runs on small lattices with gauge group SU(2).

## 2. EQUIVALENT BOSONIC THEORY

In terms of the gauge covariant forward and backward lattice derivatives, $\nabla_\mu$ and $\nabla_\mu^*$, the Wilson-Dirac operator $D$ may be written as

$$D = \tfrac{1}{2}\sum_{\mu=0}^{3}\left\{\gamma_\mu\left(\nabla_\mu^* + \nabla_\mu\right) - \nabla_\mu^*\nabla_\mu\right\}, \qquad (1)$$

where the lattice spacing has been set equal to 1 for convenience. Our starting point is the effective distribution

$$P_{\rm eff}[U] \propto \left[\det(D+m)\right]^2 \mathrm{e}^{-S_g[U]} \qquad (2)$$

of the gauge field $U$ which one obtains after integrating over the quark fields. $S_g[U]$ denotes the usual plaquette action and the bare quark mass $m$ is related to the Wilson hopping parameter $K$ through $K = (8+2m)^{-1}$.

It is easy to show that the lattice Dirac operator satisfies

$$D^\dagger = \gamma_5 D \gamma_5, \quad \|D+m\| \leq 8+m. \qquad (3)$$

The operator

$$Q = \gamma_5(D+m)/[c_M(8+m)], \quad c_M \geq 1, \qquad (4)$$

is hence hermitean and all its eigenvalues are between $-1$ and 1. In the following it will be advantageous to work with $Q$ rather than $D+m$ and so we rewrite eq.(2) in the form

$$P_{\rm eff}[U] \propto \det Q^2\, \mathrm{e}^{-S_g[U]}. \qquad (5)$$

The constant $c_M \geq 1$ will be fixed later when we discuss the simulation of the bosonic theory.

We now choose a sequence of polynomials $P(s)$ of *even* degree $n$ such that

$$\lim_{n\to\infty} P(s) = 1/s \quad \text{for all} \quad 0 < s \leq 1 \qquad (6)$$

(an explicit example is given below). From the above we then infer that

$$\det Q^2 = \lim_{n\to\infty}\left[\det P(Q^2)\right]^{-1}. \qquad (7)$$

*Talk given by M.L. at the International Symposium on Lattice Field Theory, Sept. 27–Oct. 1, 1994, Bielefeld

We may, furthermore, choose the polynomials so that their roots $z_1, z_2, \ldots, z_n$ come in complex conjugate pairs with non-zero imaginary parts. As a consequence the root factorization may be written in the form

$$P(Q^2) = \text{constant} \times \prod_{k=1}^{n} \left[(Q-\mu_k)^2 + \nu_k^2\right], \qquad (8)$$

where

$$\mu_k + i\nu_k = \sqrt{z_k}, \quad \nu_k > 0. \qquad (9)$$

After inserting this product in eq.(7) the determinant factorizes and each factor

$$\det\left[(Q-\mu_k)^2 + \nu_k^2\right]^{-1} \qquad (10)$$

can be represented as a Gaussian integral over some auxiliary bosonic field $\phi_k$.

In this way one establishes the identity

$$P_{\text{eff}}[U] = \lim_{n\to\infty} \frac{1}{\mathcal{Z}_b} \int \mathrm{D}[\phi]\mathrm{D}[\phi^\dagger]\, \mathrm{e}^{-S_b[U,\phi]}, \qquad (11)$$

where

$$S_b[U,\phi] = S_g[U] + \sum_x \sum_{k=1}^{n} \left[|(Q-\mu_k)\phi_k(x)|^2 + \nu_k^2|\phi_k(x)|^2\right] \qquad (12)$$

($\mathcal{Z}_b$ is a normalization constant independent of the gauge field).

We have thus shown that lattice QCD is a limit of a purely bosonic local theory. Note that the fields $\phi_k$ are coupled to the gauge field but not among themselves. Their action is strictly positive and the bosonic theory is, therefore, suitable for numerical simulation. Since only a relatively small number of scalar fields can be stored in memory, it is however important to make sure that the limit $n \to \infty$ is reached rapidly. In other words, the polynomial $P(s)$ must be chosen with care so as to obtain a good approximation of $1/s$ already for low values of $n$.

## 3. CHOICE OF $P(s)$

The polynomial defined below approximates the function $1/s$ with a uniform relative error in the range $\varepsilon \le s \le 1$, where $\varepsilon$ is an adjustable parameter. It is somewhat simpler than the polynomial proposed in ref.[1], but so far seems to do equally well in practice.

Let us introduce a scaled variable $u$ and an angle $\theta$ through

$$u = (s-\varepsilon)/(1-\varepsilon), \quad \cos\theta = 2u - 1. \qquad (13)$$

The Chebyshev polynomial $T_r^*(u)$ of degree $r$ is given by

$$T_r^*(u) = \cos(r\theta). \qquad (14)$$

For specified $n$ and $\varepsilon$ we now define

$$P(s) = \left[1 + \rho\, T_{n+1}^*(u)\right]/s, \qquad (15)$$

where the constant $\rho$ is chosen such that the square bracket vanishes at $s = 0$ (and so is divisible by $s$).

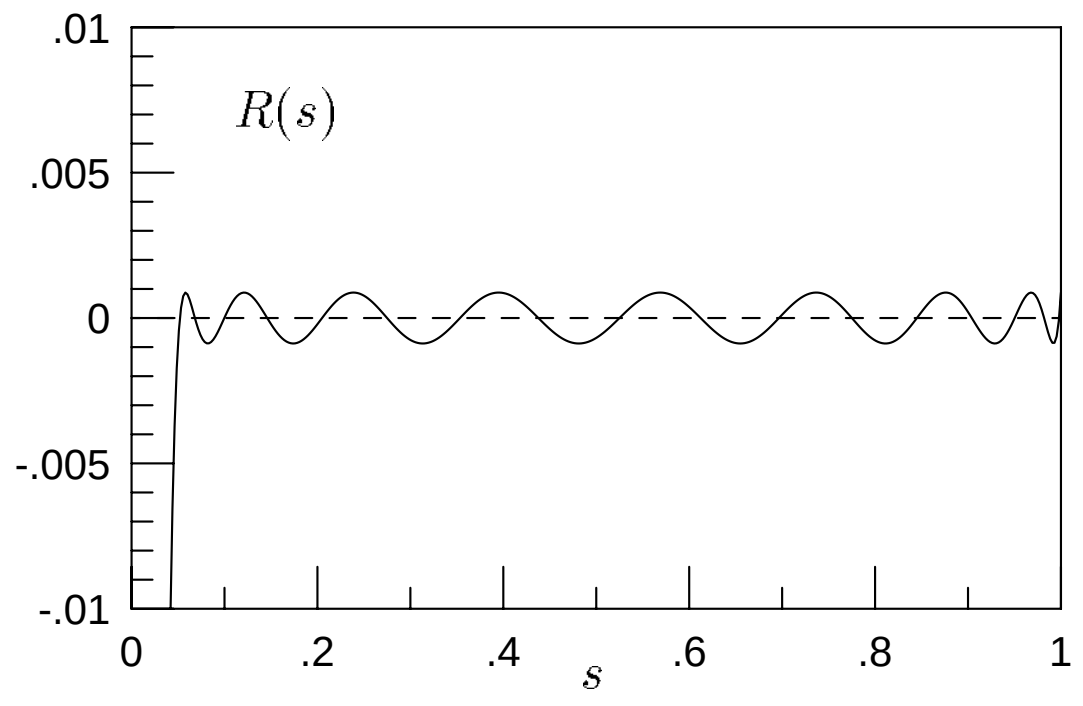

Figure 1. Relative deviation $R(s)$ of the polynomial (15) from $1/s$ for $\varepsilon = 0.05$ and $n = 16$.

In the range $\varepsilon \le s \le 1$ the relative fit error

$$R(s) = [P(s) - 1/s]\, s \qquad (16)$$

of this polynomial satisfies

$$|R(s)| \le 2\left(\frac{1-\sqrt{\varepsilon}}{1+\sqrt{\varepsilon}}\right)^{n+1}, \qquad (17)$$

i.e. the convergence is exponential with a rate roughly equal to $2\sqrt{\varepsilon}$. When $s < \varepsilon$ the polynomial continues to converge with an $s$–dependent exponential rate which approaches zero in the limit $s \to 0$. Moreover, as can be seen from fig. 1, $R(s)$ is oscillating around 0 so that the errors may be expected to partly cancel when the product over all eigenvalues of $Q^2$ is taken.

The roots of $P(s)$ are given by

$$z_k = \tfrac{1}{2}(1+\varepsilon) - \tfrac{1}{2}(1+\varepsilon)\cos\tfrac{2\pi k}{n+1} - i\sqrt{\varepsilon}\sin\tfrac{2\pi k}{n+1} \qquad (18)$$

$(k = 1, 2, \ldots, n)$. They lie on an ellipse around the spectral interval $0 \leq s \leq 1$ and satisfy $\mathrm{Im}\, z_k \neq 0$ (recall that $n$ is assumed to be even). The transformation to the bosonic theory thus works out in the way explained above.

## 4. NUMERICAL SIMULATION

The bosonic theory with action $S_b[U, \phi]$ can be simulated straightforwardly using local heatbath and over-relaxation algorithms. A few technical issues are addressed here and further details are discussed in ref.[2].

We first consider the updating of the "matter" field $\phi_k$. If all field variables are kept fixed except $\phi_k$ at point $x$, the action assumes the quadratic form

$$S_b[U, \phi] = [\phi_k(x)]^\dagger A_k \phi_k(x) + [b_k]^\dagger \phi_k(x) + [\phi_k(x)]^\dagger b_k + \text{constant}, \qquad (19)$$

where $A_k$ is an easily calculable constant positive matrix. The calculation of the (constant) vector $b_k$ requires the computation of $[Q\phi_k](x)$ and $[Q^2\phi_k](x)$. To obtain the latter one also needs to calculate $[Q\phi_k](y)$ at all neighbouring points $y$ of $x$. The subroutine which applies $Q$ to a given field at a given point must hence be called 10 times. This part of the simulation thus tends to be rather expensive.

There is, however, a more efficient way to proceed. One first computes $[Q\phi_k](z)$ at all lattice points $z$ and stores the result in an auxiliary array. After that one passes through the lattice in some order and updates the field $\phi_k$. At any given point $x$ the vector $b_k$ is obtained easily by applying $Q$ to the auxiliary field. The latter must be corrected at $z = x$ and all neighbouring points as soon as $\phi_k(x)$ has been replaced by its new value. This operation costs about as much as applying $Q$ once at a given site. In this way the total work is reduced to applying $Q$ three times per lattice point.

Concerning the updating of the gauge field, we note that the action reduces to

$$S_b[U, \phi] = \mathrm{Re}\,\mathrm{tr}\,\{U(x,\mu)F\} + \text{constant}, \qquad (20)$$

if all field variables are held fixed with the exception of the link matrix $U(x, \mu)$. The force $F$ is a sum of two contributions, the usual one from the pure gauge field action and the other from the coupling to the "matter" fields. The widely known algorithms may thus be used to refresh the gauge field. The computation of the force however requires an effort proportional to $n$, i.e. a gauge field update sweep tends to be as expensive as the updating of all fields $\phi_k$.

We finally remark that the autocorrelation times for heatbath updating of the fields $\phi_k$ with $k$ near $n/2$ tend to be rather large if the parameter $c_M$ in eq.(4) is set equal to 1 (as in ref.[1]). The probable cause for this is that the corresponding roots $z_k$ are rather close to the top $s = 1$ of the spectrum of $Q^2$. The latter can be lowered by choosing (say) $c_M = 1.1$ and the problem then disappears.

## 5. LOW-LYING EIGENVALUES OF $Q^2$

The parameter $\varepsilon$ should be chosen such that most eigenvalues $\lambda$ of $Q^2$ are in the range $\varepsilon \leq \lambda \leq 1$ where $P(\lambda)$ is a uniformly good approximation to $1/\lambda$ (if $n$ is large enough). To check whether this condition is actually met in any particular simulation, one must compute the lowest eigenvalue $\lambda_0$ of $Q^2$ for a representative sample of gauge field configurations.

There are various methods to compute the low-lying eigenvalues of large sparse matrices. A particularly safe way to obtain $\lambda_0$ is to minimize the Ritz functional

$$\mu(\psi) = \|Q\psi\|^2/\|\psi\|^2 \qquad (21)$$

in the space of all (non-zero) quark fields $\psi$, using a conjugate gradient algorithm [3–6]. Higher eigenvalues may then be computed recursively by minimizing the Ritz functional in the space orthogonal to the approximate eigenvectors already found.

It is our experience that this method works well in the case of the lattice Dirac operator. In particular, the degeneracy of the eigenvalues is correctly obtained and the attainable relative accuracy of the computed eigenvalues is completely satisfactory on machines with only single precision arithmetic, provided one employs an accurate summation method to compute scalar products of fields [7]. A rigorous bound on the final numerical accuracy may actually be deduced from the stopping criterion of the algorithm.

## 6. TEST RESULTS

We now present some results from simulations of the bosonic theory on a $6^3 \times 12$ lattice with SU(2) gauge group and bare parameters $\beta = 2.12$ and $K = 0.15$. At this point the mass $m_\rho$ of the $\rho$ meson is roughly equal to 1.28 (in lattice units), while for the pion mass $m_\pi$ one obtains $m_\pi = 0.93\, m_\rho$. The spatial size $L$ of the lattice satisfies $m_\pi L \simeq 7$ so that finite volume effects are expected to be small.

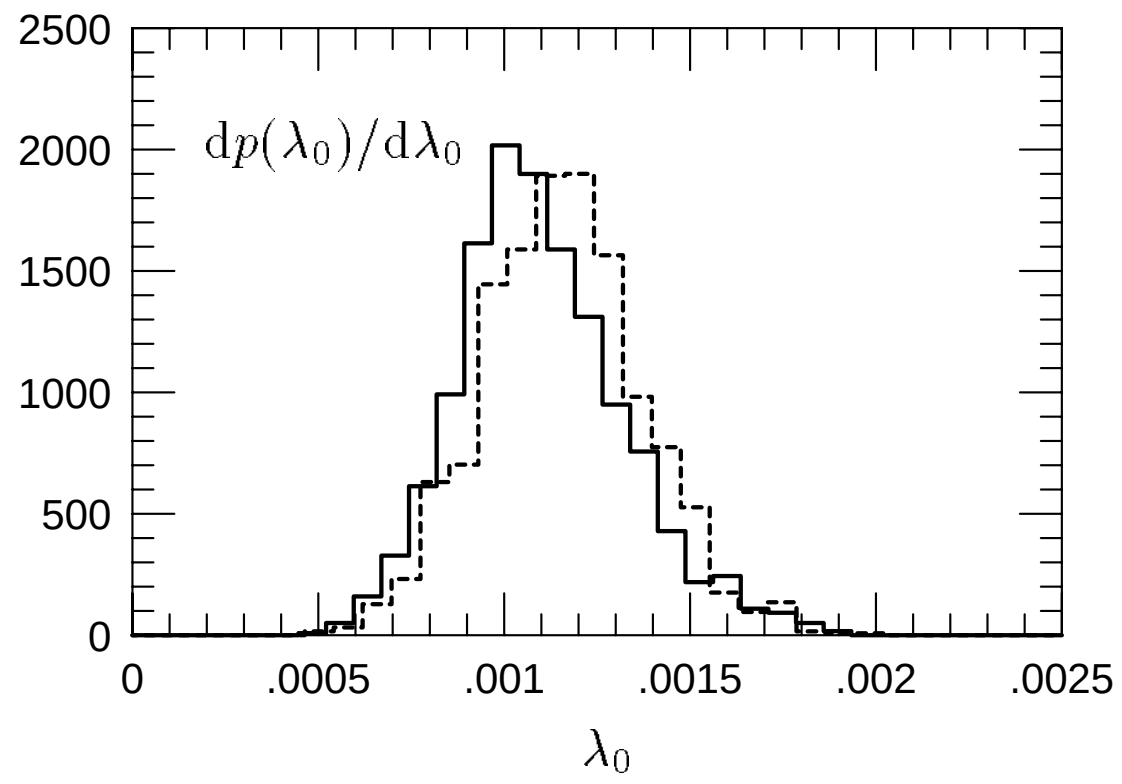

Figure 2. Probability distribution of the lowest eigenvalue $\lambda_0$ of $Q^2$. The full line is from simulations of the bosonic theory with $n = 60$ and $\varepsilon = 0.0015$, while the dashed line represents the result from the Hybrid Monte Carlo run.

In our simulations of the bosonic theory a full update cycle involved 1 heatbath plus 1 overrelaxation sweep for all matter fields $\phi_k$ followed by 1 heatbath sweep for the gauge field [8,9]. For each choice of $n$ and $\varepsilon$ we have performed at least 128'000 and sometimes up to 192'000 such cycles using the powerful APE computers at DESY-IfH. The high-quality random number generator of refs.[10,11] was used in these simulations and $c_M$ was set equal to 1.1 in all cases. For comparision we did also run a Hybrid Monte Carlo [12] program for the same lattice on the CRAY-YMP at HLRZ. Here we generated 2610 trajectories of which 1610 were used for "measurement".

The probability distribution of the lowest eigenvalue $\lambda_0$ of $Q^2$ is shown in fig. 2. Note that the distribution should be the same for all algorithms which simulate the effective gauge field distribution $P_{\rm eff}[U]$. The fact that there is an only small (and probably not significant) difference between the two curves in fig. 2 thus is a first indication that at this level of statistics one may be able to do with $n = 60$ and $\varepsilon = 0.0015$.

To further investigate this question we have made runs with $\varepsilon = 0.0004, 0.0007, 0.0015$ and values of $n$ ranging from 20 to 100. For $\varepsilon = 0.0004$ all eigenvalues $\lambda$ are always in the interval $\varepsilon \le \lambda \le 1$, while for $\varepsilon = 0.0007$ there are occasionally a few eigenvalues below $\varepsilon$. It should be emphasized that it is not unreasonable to set $\varepsilon$ to a value as large as 0.0015, since $P(\lambda)$ continues to be a good approximation to $1/\lambda$ for eigenvalues $\lambda$ substantially smaller than $\varepsilon$ (cf. sect. 7).

In fig. 3 we present our results for the plaquette expectation value. The data are plotted against the parameter

$$\delta = \max_{\varepsilon \le s \le 1} |R(s)| \tag{22}$$

which indicates how well the polynomial $P(s)$ approximates $1/s$ in the interval containing all but a few eigenvalues of $Q^2$. At fixed $\varepsilon$ and increasing $n$, $\delta$ is monotonically decreasing and eventually vanishes exponentially [eq.(17)]. In particular, the leftmost data points in fig. 3 correspond to the larger values of $n$.

The plot shows that there is a significant dependence of the average plaquette on $n$ and $\varepsilon$ when $\delta$ is greater than 5% or so. Below this level the data are consistent with a constant value and also with the result obtained from the Hybrid Monte Carlo (cf. table 1).

Table 1
Compilation of simulation results with $\delta < 0.05$

| $n$ | $\varepsilon$ | $\delta$ | $\langle$plaquette$\rangle$ | $\langle\lambda_0\rangle$ | $m_\pi$ | $m_\rho$ |
|---|---|---|---|---|---|---|
| 60 | 0.0015 | 0.0177 | 0.5802(6) | 0.00110(3) | 1.173(16) | 1.262(17) |
| 80 | 0.0007 | 0.0275 | 0.5806(8) | 0.00107(5) | 1.186(21) | 1.257(22) |
| 80 | 0.0015 | 0.0038 | 0.5801(8) | 0.00109(4) | 1.176(15) | 1.263(20) |
| 100 | 0.0004 | 0.0352 | 0.5805(9) | 0.00115(3) | 1.211(20) | 1.300(20) |
| 100 | 0.0007 | 0.0095 | 0.5814(10) | 0.00115(2) | 1.204(11) | 1.290(13) |
| 100 | 0.0015 | 0.0008 | 0.5799(7) | 0.00114(4) | 1.208(16) | 1.290(15) |
| HMC | - | - | 0.5796(6) | 0.00117(2) | - | - |

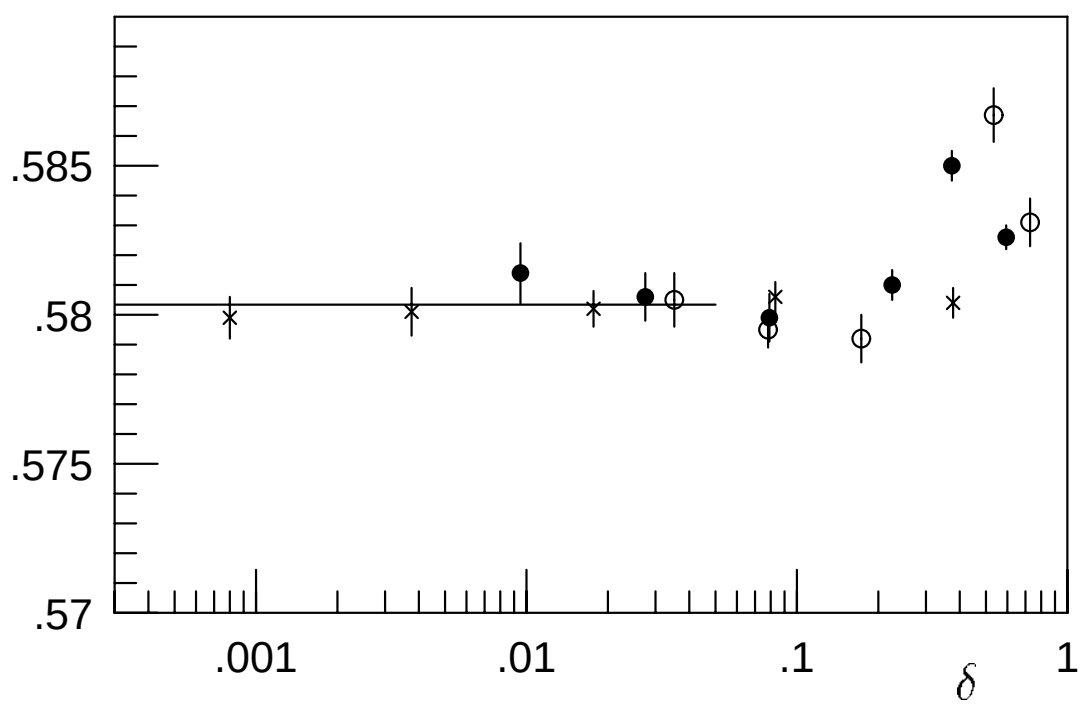

Figure 3. Plaquette expectation value plotted as a function of the fit accuracy $\delta$ [eq.(22)]. Open circles, full circles and crosses correspond to different values of $\varepsilon$ (0.0004, 0.0007, 0.0015). The line is obtained by fitting the data points with $\delta < 0.05$ by a constant.

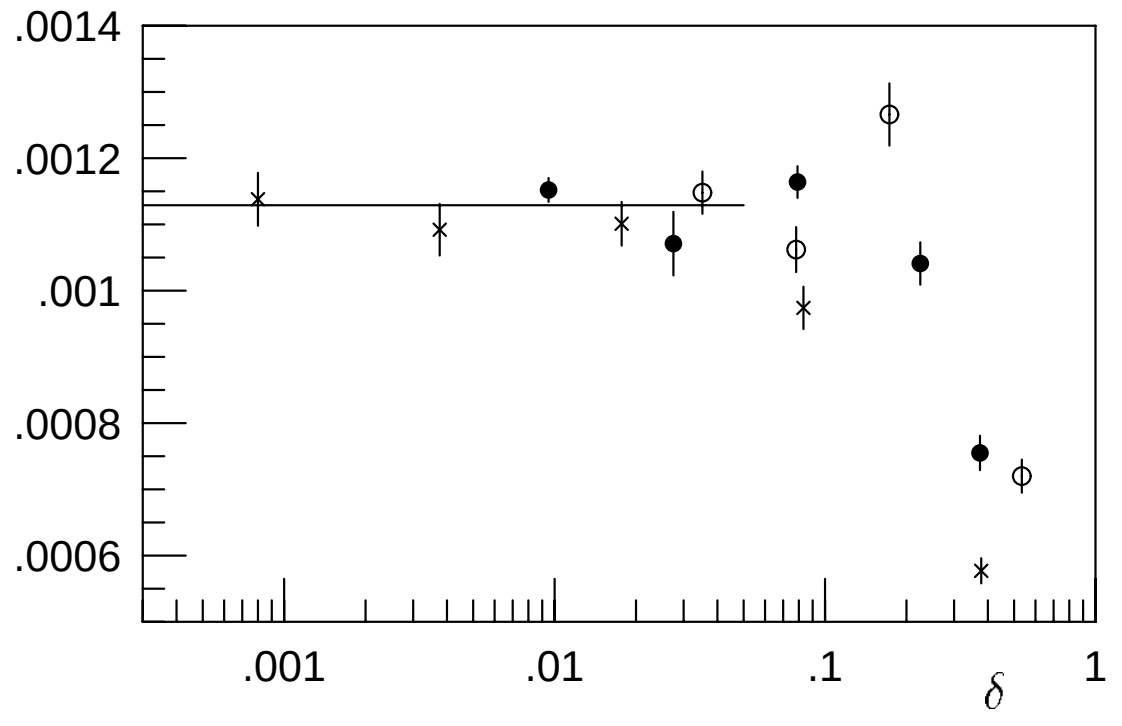

Figure 4. Same as fig. 3, but showing the expectation value of the lowest eigenvalue $\lambda_0$ of $Q^2$.

Similiar conclusions are reached when considering the expectation value of the lowest eigenvalue $\lambda_0$ of $Q^2$ (fig. 4) and the masses of the $\pi$ and $\rho$ mesons (figs. 5,6). The latter were defined through

$$\cosh m = \frac{C(T/2+1)}{C(T/2)}, \qquad (23)$$

where $C(t)$ is the appropriate correlation function of local operators at time $t$ and $T$ denotes the time size of the lattice (i.e. $T = 12$ in the present case). These masses are, therefore, not quite the same as the true masses defined through the eigenvalues of the transfer matrix. To study the properties of the new simulation algorithm the definition (23) is, however, equally useful and perhaps even better suited, since extrapolation errors are avoided. The correlation functions $C(t)$ were obtained in the conventional way, i.e. by computing quark propagators for a representative sample of gauge field configurations.

## 7. CONCLUDING REMARKS

The simulations that we have performed so far show that the new algorithm works out in the sense that correct and accurate results can be obtained with a reasonable number of boson fields. Some further consolidation by studies on larger lattices and at smaller quark masses is of course needed. One would also like to better understand how to choose the parameters $n$ and $\varepsilon$ in any given instance. For the lattice considered above one may take $n = 60$ and $\varepsilon = 0.0015$ to obtain correct results at the level of statistical precision given in table 1.

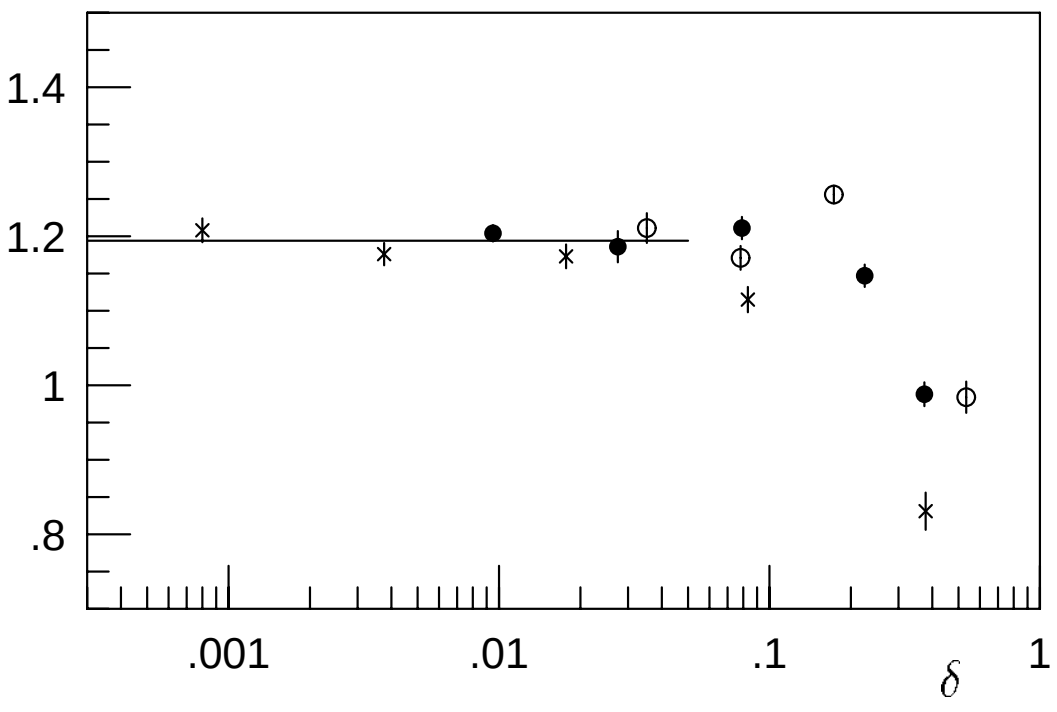

Figure 5. Results for the pion mass $m_\pi$ (cf. fig. 3).

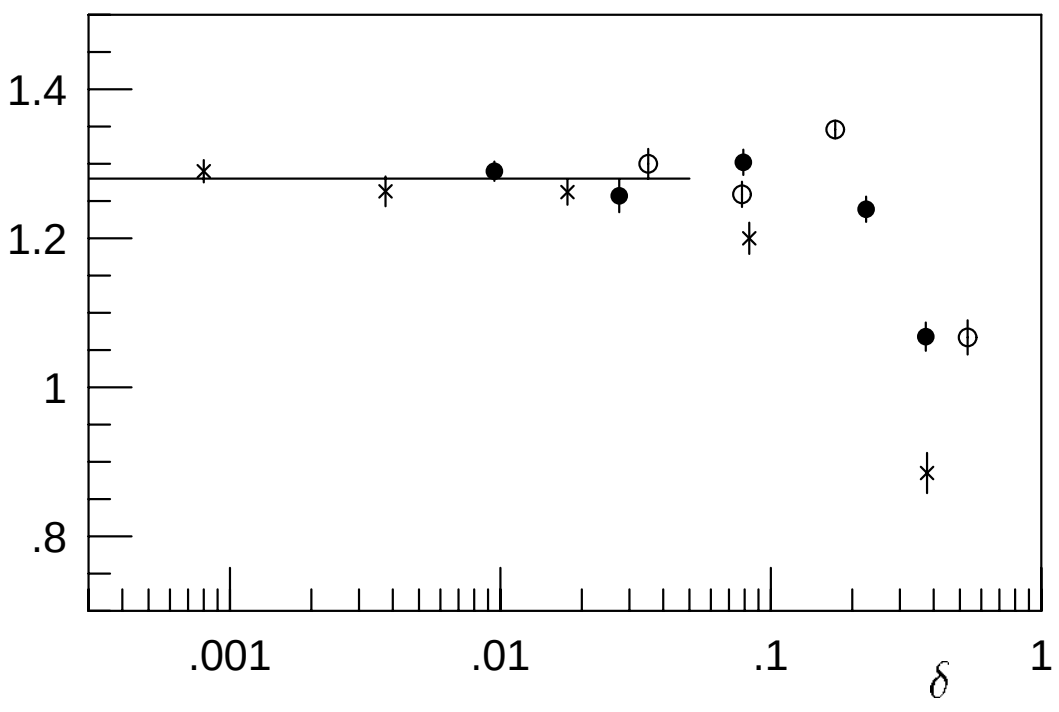

Figure 6. Results for the $\rho$ meson mass $m_\rho$ (cf. fig. 3).

It is somewhat surprising that the systematic effects (which are otherwise clearly visible in figs. 3–6) are apparently negligible for fit accuracies $\delta$ as large as a few percent. Perhaps this is due to a coherent cancellation of errors when taking the product over all eigenvalues of the quark matrix (as already discussed in sect. 3).

Another interesting observation is that when $\delta$ is smaller than about 5% the "measured" quantities did not seem to depend on the values of $\varepsilon$ we have considered. This suggests that the main effect of the quark determinant comes from the high-energy modes and that correlations between the low-lying eigenvalues in quark loops and the observed quantities are small.

Before further extensive tests are performed, we would like to install a number of improvements in the present version of the simulation program.

a. It is well-known that the condition number of the quark matrix can be significantly reduced by taking advantage of its special form with respect to the even and odd sublattices. Even-odd preconditioning can be incorporated in a rather straightforward manner in the transformation to the bosonic theory. As a result the same level of accuracy should be reached with only about half of the boson fields required without preconditioning.

b. As discussed in ref.[2] unexpectedly long autocorrelation times are observed when the bosonic theory is simulated following the lines of sect. 4. The origin of this effect is now understood and an idea has been put forward of how to accelerate the simulation algorithm [13].

c. Let $A[U]$ be any observable depending on the gauge field $U$ and $< A >$ its expectation value with respect to the true QCD distribution $P_{\rm eff}[U]$. One may then show that

$$< A >=< XA >_b \, / < X >_b, \qquad (24)$$

where $< \ldots >_b$ denotes an expectation value in the bosonic theory with action $S_b[U,\phi]$ and

$$X[U] = \det\left[1 + R(Q^2)\right]. \qquad (25)$$

Approximating $< A >$ by $< A >_b$ (as we did in the test runs) thus amounts to neglecting the correlations between $A$ and $X$. These are guaranteed to be small when $n$ is large, provided none of the eigenvalues $\lambda$ of $Q^2$ are substantially below $\varepsilon$.

In general we have

$$X[U] = \prod_{\lambda<\varepsilon} [1 + R(\lambda)] \times [1 + \mathrm{O}(\delta)]\,, \qquad (26)$$

and we may thus attempt to improve the convergence in the large $n$ limit by including $X$ in the "measurement" process, using eqs.(24) and (26). This may be particularly worthwhile in those instances, where there are just a few low-lying eigenvalues. In any case, one may in this way obtain further insight into the importance of these eigenvalues for the quantities of interest.